\documentstyle{cupconf}

\ifoldfss
\else
  \ifnfssone
    \newmathalphabet{\mathit}
      \addtoversion{normal}{\mathit}{cmr}{m}{it}
      \addtoversion{bold}{\mathit}{cmr}{bx}{it}
    \newmathalphabet{\mathcal}
      \addtoversion{normal}{\mathcal}{cmsy}{m}{n}
    \else
    \ifnfsstwo
    \fi
  \fi
\fi

\def\hexnumber#1{\ifcase#1 0\or1\or2\or3\or4\or5\or6\or7\or8\or9\or
 A\or B\or C\or D\or E\or F\fi }
\makeatletter
\ifx\CUP@mtlplain@loaded\undefined
\else
\fi
\makeatother
 \makeatletter
 \ifx\CUP@mtlplain@loaded\undefined
   \font\tenbmi=cmmib10 at 10pt
   \font\sevenbmi=cmmib10 at 7pt
   \font\fivebmi=cmmib10 at 5pt

   \newfam\bmifam
   \textfont\bmifam=\tenbmi
   \scriptfont\bmifam=\sevenbmi
   \scriptscriptfont\bmifam=\fivebmi
   
 \fi
 \makeatother

\ifnfsstwo

\fi
\ifnfssone

\fi
\ifoldfss

\fi

\mathchardef\varLambda="0103

\makeatletter
\ifx\CUP@mtlplain@loaded\undefined
\else
\fi
\makeatother

\makeatletter
\ifx\CUP@mtlplain@loaded\undefined
  \font\tenbms=cmbsy10
  \font\sevenbms=cmbsy10 at 7pt
  \font\fivebms=cmbsy10 at 5pt

  \newfam\bmsfam
  \textfont\bmsfam=\tenbms
  \scriptfont\bmsfam=\sevenbms
  \scriptscriptfont\bmsfam=\fivebms

  \edef\bsy@{\hexnumber\bmsfam}
  \mathchardef\bnabla="0\bsy@72
\fi
\makeatother

\title[Detached eclipsing binaries]{Detached eclipsing binaries as primary
distance and age indicators
\footnote{Invited talk, presented at The Extragalactic Distance Scale STScI
May Symposium, May 7 - 10, 1996, Baltimore, Maryland, USA}
}

\author[B. Paczy\'nski\/]
{B\ls O\ls H\ls D\ls A\ls N\ns P\ls A\ls C\ls Z\ls Y\ls \'N\ls S\ls K\ls I}
\affiliation{Princeton University Observatory, 124 Peyton Hall, Princeton,
NJ 08544-1001, USA}

\begin{document}
\ifnfssone
\else
  \ifnfsstwo
  \else
    \ifoldfss
      \let\mathcal\cal
      \let\mathrm\rm
      \let\mathsf\sf
    \fi
  \fi
\fi

\maketitle

\begin{abstract}
Detached eclipsing double line spectroscopic binaries offer an opportunity
to measure directly stellar parameters: mass, luminosity, radius,
as well as the distance.  The only non-trivial step is the need to determine
surface brightness of each component on the basis of something measurable,
like the color or the line ratios.  Modern model atmospheres provide a fairly
good calibration of that relation, but empirical verification is possible,
and it is needed to achieve the highest accuracy.  When this approach is
fully developed the detached eclipsing binaries should provide direct (single
step) distances with $ \sim 1\% $ accuracy to all galaxies in the Local Group.

Recent discovery of numerous detached eclipsing binaries near the main
sequence turn-off points in several globular clusters (Ka\l u\.zny et al.
1996a,b) will be followed by accurate determinations of distances,
masses and luminosities.  The empirical the mass - luminosity
relation near the main sequence turn-off point will allow accurate age
and helium determinations, and a check on the stellar evolution theory.
\end{abstract}

\firstsection
\section{Introduction}

The current status of the extragalactic distance scale and the
age estimates for globular clusters is highly unsatisfactory,
as statements like: ``Big Bang not dead yet but in decline''
may appear in one of the most prestigious scientific journals
(Maddox 1995).  The reason for the ``crisis'' is the assertion
that the best current estimates of the of age the universe, as inferred
from the adopted value of the Hubble constant, seem to be in conflict with
the best current estimates of the ages of globular clusters.
The main problem which leads to the ``crisis'' is a poor estimate
of the errors of both measurements.

Various methods used to measure distances are discussed in section 2.
The age estimates are discussed in section 3.
The discussion with emphasis on
detached, eclipsing, double line spectroscopic binaries
is presented in section 4.

\section{Distance measurements}

The more direct is the distance measurement the more trustworthy it is.
The most reliable methods are purely geometrical.  These include
trigonometric parallaxes, dynamical parallaxes, and methods based
on coherent motion of groups of objects: moving (e.g. Hyades, Schwan 1991),
rotating (e.g. NGC 4258, Miyoshi et al. 1995), or expanding (e.g. SN 1987A,
Gould 1995, and references therein).  I shall not discuss the coherent
motion methods, as they tend to be applicable to unique objects, and
it may be very difficult to verify the assumptions about the type of
motion involved.  In some cases these methods may
provide extremely accurate distances, but the errors are very difficult
to asses.

The best trigonometric parallaxes are measured with an accuracy better
than 1 mas (cf. Monet et al. 1992, Gatewood et al. 1995), and the number
of precise measurements will increase dramatically
when the Hipparcos catalog becomes available in 1997.  Unfortunately,
the accuracy of $ \sim 1\% $ can be reached only for distances smaller
than $ \sim 20 $ pc.

The range of dynamical parallaxes is considerably larger as a binary
orbit of two massive and luminous stars may be much larger than
one astronomical unit.  The linear separation between
two stars on a circular orbit can be written as
$$
A = 8.5 \times 10^{13} ~ cm ~
\left( { M_1 + M_2 \over 20 ~ M_{\odot} } \right) ^{1/3}
\left( { P_{orb} \over 3 ~ yr } \right) ^{2/3} =
5.6 ~ AU ~
\left( { M_1 + M_2 \over 20 ~ M_{\odot} } \right) ^{1/3}
\left( { P_{orb} \over 3 ~ yr } \right) ^{2/3} . 
\eqno(1)
$$
The angular separation between the two components is given as
$$
\varphi = { A \over d } = 5.4 ~ mas ~ 
\left( { M_1 + M_2 \over 20 ~ M_{\odot} } \right) ^{1/3}
\left( { P_{orb} \over 3 ~ yr } \right) ^{2/3} 
\left( { 1 ~ kpc \over d } \right) .
\eqno(2)
$$
Adopting 3 mas as the limit of resolution of the present optical 
interferometers (cf. Shao \& Colavita 1992) a massive binary can
be resolved out to a distance of $ \sim 2 kpc $. For the resolved
pairs the separation between the images of the two components can 
be measured with $ \sim 1\% $ accuracy.

The radial velocity amplitude of the same binary is given as
$$
{ K_1 + K_2  \over \sin i } = { 2 \pi A \over P } = 56 ~ km ~ s^{-1} ~
\left( { M_1 + M_2 \over 20 ~ M_{\odot} } ~
{ 3 ~ yr \over P_{orb} } \right) ^{1/3}  ,
\eqno(3)
$$
where $ i $ is the inclination of the orbit.  If this amplitude is measured
with a $ \sim 1\% $ accuracy (e.g. with the
TODCOR, a two-dimensional correlation technique, Zucker \& Mazeh 1994),
and the angular separation is known with
the same precision, then the distance to the binary can be calculated with
a $ \sim 1-2\% $ accuracy.  While accurate measurement of an angular
separation is less difficult than equally accurate measurement of a
trigonometric
parallax, the need for radial velocity measurements makes this method
somewhat more complicated.  Yet, the effective range of accurate distance
measurement is somewhat larger for the astrometric/spectroscopic binaries
than it is for trigonometric parallaxes.  Some examples
are given by Hummel \& Armstrong (1994),
Pan (1994), and Tomkin et al. (1995), and references therein.

There is a prospect to extend the range of this method with single line
spectroscopic binaries using for reference an unrelated star within
$ \sim 15'' $.
According to Colavita (1994) relative astrometry with an accuracy
reaching $ \sim 10 ~ \mu as $ can be done from the ground with a 1.5 meter
telescope for stars as faint as 16 mag at the wavelength of 2.2 $ \mu $m.
Even the Magellanic Clouds may be within reach for purely
geometrical distance determination using astrometric/spectroscopic
binaries, and the near future ground based interferometers like SUSI
(Davis 1994) and VLT (Bedding et al. 1994).

In order to reach out to globular clusters of the Milky Way, and to
galaxies of the Local Group with the existing instruments
we have to use methods which are not purely
geometrical.  The simplest of those involves detached eclipsing
double line spectroscopic binaries.  Such systems are the main source of
accurate information about stellar masses, radii, luminosities and
effective temperatures (cf. Andersen 1991, and references therein).
The procedure is very straightforward.

When two stars orbit each other in a plane which is close to the
line of sight then there are two alternate eclipses, with the
primary component eclipsing the secondary, and vice versa.
The shape of these eclipses provides the information about fractional
radii of the two stars: $ r_1 = R_1/A $ and $ r_2 = R_2/A $, where
$A$ is the separation between their centers.  The best results are
obtained when the binary is well detached, i.e. when the
quantities $ r_{1,2} $ are small.  In that case the two stars can
be assumed to be spherical, and the complications due to
mass transfer, gas streams, accretion disks, are absent.

The fractions of total light due to each component, $ L_1/L $ and
$ L_2/L $, can also be determined from the light curve analysis.
In general the so called ``third light'' may be present,
$ L_3/L = 1 - L_1/L - L_2/L > 0 $, due to
an unresolved companion.  This can be determined from an accurate
photometric light curve, as there is enough redundancy
in the shape of the two eclipses (Dreshel et al. 1989, Goecking et al. 1994,
Gatewood et al. 1995).  The light curve analysis also provides
the values of orbital period, inclination, and eccentricity: P, i, e.

The radial velocity curves provide information about
the two amplitudes $ K_1 $ and $ K_2 $, and the orbital
eccentricity.  The latter offers a consistency check
with the photometry.  Given the radial velocity curves, the orbital
period and inclination, the size of the orbit can be calculated.
Finally, the linear radii of the two stars, $ R_1 $ and $ R_2 $, can
be calculated too.

So far the analysis was purely geometrical.  Given good enough photon
statistics the errors of all quantities can be reduced below 1\% level, as
demonstrated by many investigators (cf. Andersen 1991).  The next step is 
the only one which involves physics: the surface brightness of the two 
components must be estimated on the basis of the observed colors, spectral
line ratios, or by whatever means.  
Let these two quantities be $ F_1 $ and $ F_2 $.
Given those, the absolute luminosity
of each star in the observed photometric band can be calculated as:
$$
L_1 = 4 \pi R_1^2 F_1 , \hskip 2.0cm
L_2 = 4 \pi R_2^2 F_2 .
\eqno(4)
$$
The flux of radiation from each star at the telescope is directly measured
as $ F_{1,tel} $ and $ F_{2,tel} $.  Therefore, the distance
between each star and the telescope can be calculated:
$$
d_1 = \left( { L_1 \over 4 \pi F_{1,tel} } \right) ^{1/2} =
\left( { F_1 \over F_{1,tel} } \right) ^{1/2} R_1 ,  \hskip 0.5cm
d_2 = \left( { L_2 \over 4 \pi F_{2,tel} } \right) ^{1/2} =
\left( { F_2 \over F_{2,tel} } \right) ^{1/2} R_2 .
\eqno(5)
$$
In these equations the two stellar radii follow directly from the
analysis of the photometric and radial velocity curves.
Naturally, the two values for the distance, $ d_1 $ and $ d_2 $,
should be the same, within
errors.  The procedure can be repeated for many photometric bands, 
and all distance determinations should be the same, within errors.
For any well observed binary a considerable
redundancy is provided by this method.

While in principal any photometric band should yield the same distance,
some bands are better than others.  The two main difficulties are:
the calibration of the surface brightness scale (related to the effective
temperature scale, cf. B\"ohm-Vitense 1981),
and the correction for interstellar reddening.
Both are minimized in the infrared, where the surface brightness is
proportional to the first power of effective temperature, and the interstellar
reddening is the smallest.  Naturally, the estimate of the effective
temperature and the reddening should be done in the part of the spectrum
which is most sensitive to these two quantities, i.e. either in the ultraviolet
or in the blue part of the spectrum.

The first case of a similar procedure was published many decades ago by
Stebbins (1910).  At that time the trigonometric distance, the photometric
orbit and the radial velocity curves were well measured for the nearby
binary $ \beta $ Aurigae, and Stebbins used those measurements to determine
the surface brightness of the two identical components.  Today the relation
between the spectrum and surface brightness is reasonably well known from
theoretical models (cf. Kudritzki \& Hummer 1990, Kurucz 1992,
Buser \& Kurucz 1992,
Sellmaier et al. 1993).  An example of practical application of the 
eclipsing binary method is provided by Milone, Stagg \& Kurucz
(1992), who determined the distance to the binary system AI Phoenicis to
be $ 170 \pm 9 $ pc.  

The theoretical relation between the stellar spectrum (or color) and
surface brightness should
be verified empirically.  The best opportunity is provided by the stars for
which angular diameters were measured (cf. McAlister 1985,
Shao \& Colavita 1992, and references therein).  It is my impression
that current accuracy is better than $ \sim 5\% $ for the value of
effective temperature over a broad range of spectral types, and there 
is good prospect for making it much more accurate.

Somewhat similar but far more complicated, and also far more popular, are
Baade-Wesselink and Barnes-Evans methods (cf. Welch 1994, Feast 1995, Walker
1995, and references therein) as applied to pulsating stars.  One of their
weak points is the same as the only weak point of the eclipsing binary method:
it is necessary to infer
stellar surface brightness from the spectrum (color or line ratios).
However, the B-W and B-E methods have additional weaknesses of their
own: the spectrum of a pulsating star is never strictly in a hydrostatic
equilibrium, the measured radial velocity has to be converted to the
radial velocity of stellar expansion or contraction, the optical depths
at which the lines and the continuum are formed are not fixed with
respect to matter.  

A related Expanding Photosphere Method is used to determine distances to
Type II supernovae (cf. Schmidt, Kirshner \& Eastman 1992, and references
therein).  It suffers from all the ailments of the B-W and B-E methods.
In spite of all the problems the actual results for cepheids and for 
supernovae are very encouraging, as presented by Kirshner (1996) and
Tanvir (1996).   Therefore, there is every reason to expect that the
simpler and more direct method of measuring distances to detached eclipsing
binaries will be even more successful, as pointed out by
Guinan, Bradstreet, \& DeWarf (1995), Gimenez et al. (1995),
Bradstreet et al. (1995), and Paczy\'nski (1996a).

A preliminary estimate of the
LMC distance modulus using detached eclipsing binaries
is $ (m-M)_{LMC} = 18.6 \pm 0.2 $ (cf. Guinan et al.
1995).  This accuracy is not very impressive, but it will be
greatly improved in the next few years.  There are many 
developments which lead to this optimism.

Well detached eclipsing binaries are difficult to find because they
are so well detached, and hence their eclipses are very narrow.
With the massive CCD photometry developed in the last few years
as a by-product of microlensing searches (cf. Paczy\'nski 1996b,
and references therein) tens of thousands of variable stars were discovered.
The first catalogs of eclipsing binaries were already published
(Grison et al. 1994, Ka\l u\.zny et al. 1996a,b,
Udalski et al. 1994, 1995a,b, 1996).  The large number of objects makes it
possible to select the most promising systems for the more detailed
studies.  The best binaries have narrow and deep primary and secondary 
eclipses, indicating that the two stars are pretty similar, and their radii
are much smaller than their separation.  Also, no variability should
be present outside the eclipses, indicating that the proximity effects
are small, and there are no major spots on the stellar surfaces.
The search for such systems can be readily done with a 1-meter, or even
smaller telescope, but a large area CCD camera and dozens
of observing nights are necessary, as demonstrated by the microlensing
searches..

For a subset of promising systems the accurate photometry should be 
done on a medium size telescope, so that photon noise is 
eliminated as a significant source of errors.  When a photometric
orbit is obtained it should become clear if the system is indeed
simple, i.e. can the shapes of both eclipses be accurately fitted,
and can the ``third light'' be accurately determined.  Presumably,
only some of the systems with accurate photometry will turn out
to be suitable, i.e. will provide very accurate values for the
fractional radii of the two components, as well as accurate measurement
of the fraction of light from each component.

The third stage is the determination of the best possible radial velocity
curves and the spectral measurement.  The largest telescopes are required
for this step to provide a very high S/N.  Accurate 
measurements of radial velocities of both components are now possible
with the recently developed TODCOR method (Zucker \& Mazeh 1994, Metcalfe
et al. 1995).  Relative temperatures for cool stars can be determined
spectroscopically to 30 K accuracy (Sasselov \& Lester 1990), and
the prospect to have a very accurate absolute calibration for early type 
stars are also good (E. Fitzpatrick, private communication).
Note, that surface gravity follows directly from the
values of masses and radii, i.e. it is measured directly for
the components of eclipsing binaries.  Therefore,
the temperature and the chemical composition are the only
parameters which have to be inferred from the spectra (or colors).

While the determination of interstellar reddening is best done in the
blue or even UV part of the spectrum, the infrared K-band
measurements are the least affected by the reddening and effective
temperature errors (cf. Kelly, Rieke \& Campbell 1994, and
references therein).

\section{Age determinations}

The conventional age estimates are based on a comparison between the 
observed and theoretical color-magnitude diagrams for stars in globular 
clusters (cf. Chaboyer, 1995, and references therein).
The critical region is near the main sequence turn off point (TOP),
where stars are similar to the sun.  There are two problems with
this method: the dependence on the distance and on the ``mixing
length theory''.

The age of a cluster is inversely proportional
to the TOP luminosity, $ t_{cluster} \sim L_{TOP}^{-1} $
(cf. Bergbush \& VandenBerg 1992, Bertelli et al. 1994,
or any other set of theoretical isochrones).
For a given observed flux of radiation, $ F_{tel} $
the luminosity is proportional 
to the distance square, $ L_{TOP} \sim d_{cluster}^2 $.  Therefore, the 
age estimate is inversely proportional to the distance estimate,
$$
t_{cluster} \sim d_{cluster}^{-2} .  \eqno(6)
$$
If the distance is known to 10\% accuracy then the age
is known to 20\%.  It is clear that for the ages to be measured
accurately, the distances have to be known even more accurately.
This is not a fundamental problem, but it has to be faced and solved.

There are various attempts to circumvent the distance uncertainty.
For example, one may use the magnitude difference between the horizontal
branch and the TOP.  If that was to be taken seriously, and if $ \sim 10\% $
errors in the ages were believable, then one could use the eq. (1)
to measure distances to clusters with $ \sim 5\% $ accuracy.  I am not aware
of this approach ever used to measure distances.  Apparently, the
accuracy of age estimates is not taken seriously.

The second, more fundamental problem with the color-magnitude diagram
is its dependence on the ``mixing length theory'' of
sub-photospheric convection.  The radii of the models, and their effective
temperatures, and hence colors, depend on the assumed efficiency of
convection through the choice of $ \alpha $ parameter.  In practice the value 
of $ \alpha $ has to be obtained empirically.  There is no reason for this
parameter to be constant throughout the color - magnitude diagram.
In fact, theoretical isochrones which
fit the TOP for some adopted value of $ \alpha $ are systematically
different from the observations in the subgiant region (e.g. Bergbush \&
VandenBerg 1992).  Various
new, more sophisticated models of convection are being developed (Caloi
et al. 1996, Canuto 1996,
Canuto et al. 1996, Spruit 1996), and it is clear that
the shape of isochrones on the color-magnitude diagram depends on the
formulation of the theory.  As long as a quantitative theory of convection
is not available the use of color-magnitude diagrams is subject to
errors which are very difficult, perhaps impossible to estimate.

The distance determination to globular clusters can be made more accurate
using detached eclipsing binaries.  Five such binaries were recently discovered
near the main sequence turn-off point in Omega Cen (Ka\l u\.zny et al. 
1996a,b), and five more detached eclipsing binaries were found in M4 
(Ka\l u\.zny \& Thompson 1996).  The search will no dobt be extended
to many more clusters.  Following the procedure
described in the previous section the distances to those globular
clusters will become known with a few percent accuracy, perhaps even
$ \sim 1\% $ accuracy will be reached.  This will take care of one major
contribution to the age uncertainty.

Once the masses of binary components are determined, and their bolometric
luminosities are measured, it will be possible to use the empirical
mass-luminosity relation to infer the age and helium content by comparing 
the observed stars with
the  standard stellar models, like those of Bergbush \& VandenBerg (1992)
and Bertelli et al. (1994).  For a given chemical composition the mass of
TOP stars varies with age according to $ M_{TOP} \sim t_{cluster}^{-1/3.7} $.
The mass determination is based on the radial velocity curves, with
$ M \sim K^3 $, where $ K $ is the radial velocity amplitude.  Therefore,
the age estimate is given as
$$
t_{cluster} \sim M_{TOP}^{-3.7} \sim K^{11},
\eqno(7)
$$
a very steep relation, indeed.  However, the value of $ K_{1,2} $ for
a binary system CM Draconis has been recently measured with an accuracy
of 0.2\% by Metcalfe et al. (1996) using the TODCOR method (Zucker \& Mazeh
1994).  If a comparable precision is reached for binaries in
globular clusters then the errors in radial velocity measurements will
contribute only $ \sim 2\% $ to the age error.

Other parameters being equal the luminosity of the TOP stars is
proportional to a high power of the mean molecular weight, $ L \sim \mu ^7 $,
and that is strongly affected by the helium content.  Therefore, the
mass-luminosity relation of the TOP stars can be used to determine not
only their age, but also their helium content, assuming that the ratio Z/X
is known spectroscopically.  Note that stellar luminosity 
is practically independent of the sub-photospheric convection
(Paczy\'nski 1984).

\section{Discussion}

The existing technology: photometry, spectroscopy, empirically
calibrated spectrum - surface brightness relation, offer a prospect
of using detached eclipsing spectroscopic binaries for very accurate
distance measurements to globular clusters, to the Magellanic Clouds, and
even to M31 and M33.  There is no fundamental reason why the precision
as high as $ \sim 1\% $ could not be achieved, with large telescopes 
providing a large number of photons and a high S/N.  The weakest link is the
relation between the observed spectrum (colors, line ratios, Balmer jump)
and the surface brightness.  The best prospect is to use model atmospheres
to interpolate between the few most accurate empirical calibrators, which
are provided by stellar angular diameters measured with
optical interferometers.  The errors of estimates of the surface
brightness and the interstellar extinction can be minimized with
K-band photometry.

This is not a new method.  The principle goes back to Stebbins (1910).
The method was successfully applied to some bright binaries, like AI 
Phoenicis (Milone et al. 1992).  It was used to estimate the LMC distance
(Guinan et al. 1995).  However, it is little known, and it is not referenced
in major papers, like the review of fundamental parameters of 
detached binaries (Andersen 1991), or the review of
distance determinations to nearby galaxies
(Huterer et al. 1995).  Nevertheless, this method
may be the best for distance measurements
to globular clusters and to galaxies of the Local Group.
It is likely to be superseded only by some future interferometric measurements.

The single most uncertain element of the method is the relation between
the observable spectra (colors, line ratios, Balmer jump) and the surface
brightness of binary components.
The distances to nearby galaxies 
will be measured using the brightest stars which
do not have complications caused by stellar winds, i.e. 
the early B type stars on the main sequence.  Distances to globular
clusters will be  measured using main sequence (TOP)
stars of low and moderate metalicity, i.e. the spectral types F and G.
Therefore, the calibration of two ranges of spectral types is
most important, the early B type and F-G types.

The masses and absolute luminosities of the TOP stars 
in globular clusters can be used to determine the ages and helium
contents of those clusters if the values of mass and luminosity are available
for at least two stars in a given cluster.  If more than 2 different pairs
of data points are available then the shape of empirical 
mass-luminosity relation
can be used to check stellar models. 
The poorly understood sub-photospheric convection has practically
no effect on the mass-luminosity relation (cf. Paczy\'nski 1984).

In order to take full advantage of the future measurements of
masses and bolometric luminosities it is necessary
to expand the network of evolutionary
tracks to cover both dimensions in the (Y,Z) plane.  The models
of Bergbush \& VandenBerg (1992) and Bertelli et al. (1994)
cover just a line in that plane.

This work was supported with the NSF grants AST-9216494 and AST-9528096.

{}


\begin{thebibliography}{} 

\bibitem[]{}
Andersen, J. 1991, A\&AR, 3, 91

\bibitem[]{}
Bedding, T. R. et al. 1994,
IAU Symposium 158: ``Very High Resolution Imaging'' (Eds.: J. G. Robertson
\& W. J. Tango, Kluver Academic Publishers. Dordrecht), p. 143

\bibitem[]{}
Bergbush, P. A., \& VandenBerg, D. A. 1992, ApJS, 81, 163

\bibitem[]{}
Bertelli, G. et al. 1994, A\&AS, 106, 275

\bibitem[]{}
B\"ohm-Vitense, E. 1981, ARA\&A, 19, 295

\bibitem[]{}
Bradstreet, D. H. et al. 1995, BAAS, 187, 43.20

\bibitem[]{}
Buser, R., \& Kurucz, R. L. 1992, A\&A, 264, 557

\bibitem[]{}
Colavita, M. M. 1994,
IAU Symposium 158: ``Very High Resolution Imaging'' (Eds.: J. G. Robertson
\& W. J. Tango, Kluver Academic Publishers. Dordrecht), p. 469

\bibitem[]{}
Caloi, V. et al. 1996, this Symposium

\bibitem[]{}
Canuto, V. 1996, ApJ, 467, 385

\bibitem[]{}
Canuto, V. et al. 1996, preprint

\bibitem[]{}
Chaboyer, B. 1995, ApJ, 444, L9

\bibitem[]{}
Davis, J. 1994,
IAU Symposium 158: ``Very High Resolution Imaging'' (Eds.: J. G. Robertson
\& W. J. Tango, Kluver Academic Publishers. Dordrecht), p. 135

\bibitem[]{}
Dreshel, H. et al. 1989, A\&A, 221, 49

\bibitem[]{}
Feast, M. 1995,
ASP Conf. Ser. Vol. 83: ``Astrophysical Applications of Stellar Pulsations''
(Eds.: R. S. Stobie \& P. A. Whitelock, BookCrafters, Inc.), p. 209

\bibitem[]{}
Gatewood, G. et al. 1995, AJ, 109, 434

\bibitem[]{}
Gimenez, A. et al. 1995, Experimental Astron., 5, 181

\bibitem[]{}
Goecking, K.-D.  et al. 1994, A\&A, 289, 827

\bibitem[]{}
Gould, A. 1995, ApJ, 452, 189

\bibitem[]{}
Grison, P. et al. 1994, A\&AS, 109, 447

\bibitem[]{}
Guinan, E. F., Bradstreet, D. H., \& DeWarf, L. E. 1995, in ``The Origins,
Evolution, and Destinies of Binary Stars in Clusters'', (Eds.: E. F. Milone
\& J. C-Mermilliod), ASP Conf. Ser. 90, p. 196

\bibitem[]{}
Hummel, C. A., \& Armstrong, J. T. 1994,
IAU Symposium 158: ``Very High Resolution Imaging'' (Eds.: J. G. Robertson
\& W. J. Tango, Kluver Academic Publishers. Dordrecht), p. 410

\bibitem[]{}
Huterer, D., Sasselov, D. D., \& Schechter, P. L. 1995, AJ, 110, 2705

\bibitem[]{}
Ka\l u\.zny, J. et al. 1996a, A\&A, submitted = preprint astro-ph/9601053

\bibitem[]{}
Ka\l u\.zny, J. et al. 1996b, A\&A, submitted = preprint astro-ph/9604027

\bibitem[]{}
Ka\l u\.zny, J., \& Thompson, I. 1996, in preparation

\bibitem[]{}
Kelly, D. M., Rieke, G. H., \& Campbell, B. 1994, ApJ, 425, 231

\bibitem[]{}
Kirshner, R. 1996, this Symposium

\bibitem[]{}
Kudritzki, R. P., \& Hummer, D. G. 1990, ARA\&A, 28, 303

\bibitem[]{}
Kurucz, R. L. 1992, in IAU Symp. 149: ``The Stellar Populations in Galaxies'',
(Eds.: B. Barbuy \& A. Renzini, Kluwer Acad. Publ., Dordrecht), p. 225

\bibitem[]{}
McAlister, H. A. 1985, ARA\&A, 23, 59

\bibitem[]{}
Maddox, J. Nature, 377, 99

\bibitem[]{}
Metcalfe, T. S. et al. 1996, ApJ, 456, 356

\bibitem[]{}
Milone, E. F., Stagg, C. R., \& Kurucz, R. L. 1992, ApJS, 79, 123

\bibitem[]{}
Miyoshi, M. et al. 1995, Nature, 373, 127

\bibitem[]{}
Monet, D. G. et al. 1992, AJ, 103, 638

\bibitem[]{}
Paczy\'nski, B. 1984, ApJ, 284, 670

\bibitem[]{}
Paczy\'nski, B. 1996a, in ``Astrophysical Applications of Gravitational
Lensing'', (Eds.: C. S. Kochanek \& J. N. Hewitt), IAU Symp. 173, p. 199

\bibitem[]{}
Paczy\'nski, B. 1996b, ARA\&A, Vol. 34, in press = astro-ph/9604011

\bibitem[]{}
Pan, X. 1994,
IAU Symposium 158: ``Very High Resolution Imaging'' (Eds.: J. G. Robertson
\& W. J. Tango, Kluver Academic Publishers. Dordrecht), p. 413

\bibitem[]{}
Sasselov, D. D., \& Lester, J. B. 1990, ApJ, 360, 227

\bibitem[]{}
Schmidt, B. P., Kirshner, R. P., \& Eastman, R. G. 1992, ApJ, 395, 366

\bibitem[]{}
Schwan, H. 1991, A\&A, 243, 386

\bibitem[]{}
Sellmaier, F. et al. 1993, A\&A, 273, 533

\bibitem[]{}
Shao, M., \& Colavita, M. M. 1992, ARA\&A, 30, 457

\bibitem[]{}
Spruit, H. 1996, preprint: astro-ph/9605020

\bibitem[]{}
Stebbins, J. 1910, ApJ, 32, 185

\bibitem[]{}
Tanvir, N. 1996, this Symposium

\bibitem[]{}
Tomkin, J., Pan, X., \& McCarthy, K. 1995, AJ, 109, 780

\bibitem[]{}
Udalski, A. et al. 1994, AcA, 44, 317

\bibitem[]{}
Udalski, A. et al. 1995a, AcA, 45, 1

\bibitem[]{}
Udalski, A. et al. 1995b, AcA, 45, 433

\bibitem[]{}
Udalski, A. et al. 1996, AcA, 46, 51

\bibitem[]{}
Walker, A. R. 1995, 
ASP Conf. Ser. Vol. 83: ``Astrophysical Applications of Stellar Pulsations''
(Eds.: R. S. Stobie \& P. A. Whitelock, BookCrafters, Inc.), p. 198

\bibitem[]{}
Welch, D. L. 1994, AJ, 108, 1421

\bibitem[]{}
Zucker, S., \& Mazeh, T, 1994, ApJ, 420, 806

\end{thebibliography}
\end{document}